\documentclass{raa}
\usepackage{graphicx,times}
\usepackage{natbib}
\usepackage{amssymb,amsmath}
\usepackage{esint}
\bibpunct{(}{)}{;}{a}{}{,}

\usepackage[a4paper=true,dvipdfm=true,pagebackref=true]{hyperref}
\hypersetup{pdftitle = The title of my PDF, pdfauthor = My name, pdfsubject= The subject, pdfkeywords = keyword1 keyword2 keyword3}
\hypersetup{colorlinks = true, linkcolor = green, anchorcolor = red, citecolor = blue, filecolor = red, pagecolor = red, urlcolor = red}

\begin{document}

   \title{The Induced Electric Field Distribution in Solar Atmosphere
}

 \volnopage{ {\bf 2012} Vol.\ {\bf X} No. {\bf XX}, 000--000}
   \setcounter{page}{1}

   \author{Rong Chen\inst{1,2}, Zhiliang Yang\inst{1,2}, Yuanyong Deng\inst{2}
   }

   \institute{ Department of Astronomy, Beijing Normal University, Beijing 100875,
China; {\it zlyang@bnu.edu.cn}\\
 \and
 Key Laboratory of Solar Activity, National Astronomical Observatories, Chinese Academy of Sciences, Beijing 100012
 \vs \no}

\abstract{A method of calculating induced electric field is presented in this paper. Induced electric field in solar atmosphere is derived by the time variation of magnetic field when the charged particle accumulation is neglected. In order to get the spatial distribution of magnetic field, several extrapolation methods are introduced. With observational data from Helioseismic and Magnetic Imager (HMI) aboard the NASA's Solar Dynamics Observatory (SDO) on May 20th, 2010, we extrapolate the magnetic field to the upper atmosphere from the photosphere. By calculating the time variation of magnetic field, we can get the induced electric field. The derived induced electric field can reach a value of $10^2$ V/cm and the average electric field has a maximum point at the layer of 360 km above the photosphere. The Monte Carlo statistics method is used to compute the triple integration of induced electric field.
}

   \authorrunning{R. Chen et al. }            
   \titlerunning{The induced electric field distribution in solar atmosphere}  
   \maketitle

%
\section{Introduction}           
\label{sect:intro}

Electric field in solar atmosphere plays an important role in heating plasma, and accelerating and transporting charged particles (\citealt{Priest+Forbes+2000}). At the same time, its distribution provides rich information of solar flare, as well as other dynamic solar activities. The simultaneous determination of electric and magnetic  field vectors make the estimation of Poynting flux of electromagnetic energy entering the corona and the flux of relative magnetic helicity available. However, the determination of electric field is quite harder and less than that of magnetic field in the solar atmosphere.

Although the solar activities are dominated by magnetic filed, and much progress was made in this aspect in the past decades, there are still many points out of our understanding, such as the physical mechanisms of flares and filament eruptions. It is time to figure out that electric field, its magnitude, geometry, time-dependence and especially spatial distribution do provide us powerful tools to probe in solar activities where particle accelerating and energy release are believed to occur.

\cite{Wien+1916} is the first author who pointed out the possibility of measuring electric field of solar plasma and measured the motional electric field using Stark effect. Around 1980s, some attempts to measure the electric field with helium and silicon spectra had been made, which suggested an electric field of 700 V/cm (\citealt{Davis+1977})
and 300 V/cm (\citealt{Jordan+etal+1980}) respectively. Even so, \cite{Moran+Foukal+1991} pointed out that direct measurement of the electric field by Stark effect is hard to operate with the low sensitivity.

Because there was not an  efficient and reliable way to get the direct electric field of solar atmosphere, researchers thought out many indirect methods, which either explicitly or implicitly invoke the ideal MHD assumption, $\vec{E} = -\vec{v} \times \vec{B}/c$. Tracking method and inductive method are two classes of such techniques.

According to the ideal MHD equation above, the magnetic field is known from the vector magnetogram, so both of two classes have focused on determining the velocity vector. The tracking method, developed by \cite{November+Simon+1988}, compute velocity through a cross-correlation	function that depends on the shift of feature points between two images. Although tracking methods are simple and robust, they also suffer from some shortcomings, for example, this technique is actually two-dimensional without the vertical component. Inductive methods , first developed by \cite{Kusano+etal+2002}, improves the result of tracking methods with a solution to the vertical component from the magnetic induction equation and derives a three-components velocity vector from a sequence of vector magnetograms. Since the seminal work of \cite{Kusano+etal+2002}, several techniques have been developed to determine velocity from vector magnetograms, and \cite{Welsch+etal+2007} provided detailed tests and comparisons of these techniques.

\cite{Poletto+Kopp+1986} derived the maximum electric field of 2 V/cm in a large two-ribbon flare using the reconnection theory of \cite{Priest+Forbes+2000} where they used a very simple relationship between the electric filed along the current sheet and the observable velocity and magnetic field. With the similar theory, \cite{Wang+etal+2003} discovered two stages of electric evolution in another two-ribbon flare: electric field remains 1 V/cm averaged over 20 minutes as the first stage, and are followed by 0.1 V/cm in the next 2 hours. \cite{Qiu+etal+2002} worked on an impulsive flare with high cadence H$\alpha$ observations at Big Bear Solar Observatory(BBSO) and estimated the maximum electric field to be 90 V/cm.

Recently, \cite{Fisher+etal+2010} pointed out a way to compute electric field from a sequence of the vector magnetograms by using the Faraday's law and showed that it is possible to derive electric field whose curl is the time derivatives of three-components of $\vec{B}$. The main problem for these authors is the non-unique solution of Faraday's law.

In this paper, our goal is to present a method to indirectly compute the three-components induced electric field distribution in solar atmosphere through evolution of vector magnetic field. We first extrapolate the magnetic field from photosphere to corona, and then calculate the induced electric field from the time variation of magnetic field.

In solar plasma, there are three kinds of electric field: static electric field, induced electric field and motional electric field. Here we only focus on induced electric field which is caused by the change of magnetic field. The static electric field which is triggered by existence of charged particles accumulation and the motion of plasma in the direction perpendicular to magnetic field are not considered due to the screening of plasma in the solar atmosphere.

The paper is structured as follows. In section 2 we extrapolate vector magnetic field distribution from  magnetogram; in section 3 we do the computing of electric field from the extrapolated magnetic field; and next in section 4 we give an examination by use of the observational vector magnetograms from HMI/SDO. Conclusion and discussion are given in section 5.

\section{Extrapolation of Magnetic Field}

At present, although many attempts had been done to estimate the coronal magnetic field (\citealt{House+1977}; \citealt{Arnaud+Newkirk+1987}; \citealt{Judge+1998}; \citealt{Judge+etal++2001}), the reliable information about magnetic field is available only for the photospheric level. Just similar to the electric filed, we currently and in the near future have to face the fact that the direct measurements of magnetic field in the global solar atmosphere are still unavailable.

There are many methods to extrapolate magnetic field from the photospheric magnetic field under the assumption that the magnetic field is nearly force-free. Force-free magnetic field of solar atmosphere must satisfy the following equations:
\begin{equation}
  \vec{j} \times \vec{B} = 0,
\label{eq:LebsequeI}
\end{equation}

\begin{equation}
  \nabla \times \vec{B} = \alpha \vec{B},
\label{eq:LebsequeI}
\end{equation}

\begin{equation}
  \nabla \cdot \vec{B} = 0,
\label{eq:LebsequeI}
\end{equation}
where $\alpha$ is a scalar function of position and time. The above equations imply nonexistence of Lorentz force and $\alpha$ being a constant along the magnetic field line. The equations represent a potential field if $\alpha$ = 0,  a current-carrying linear force-free(LFF) field if $\alpha$ = constant,  and a general nonlinear force-free field(NLFF) field if $\alpha = f(\vec{r})$ respectively.

The extrapolations of potential and linear LFF field are maturely developed. Potential field and LFF field can be determined directly from the line-of-sight (LOS) component of magnetic field(e.g. MDI/SOHO) as input, and $\alpha$ has to be computed in LFF field from some additional data(\citealt{Chiu+Hilton+1977}; \citealt{Seehafer+1978}; \citealt{Alissandrakis+1981}; \citealt{Gary+1989}).

For the NLFF field, several methods have already been pointed out: Grad-Rubin method (\citealt{Sakurai+1981}, MHD relaxation method (\citealt{Chodura+Schlueter+1981} \citealt{Roumeliotis+1996}) and the optimization method (\citealt{Wheatland+etal+2000}). The last one will be used in this paper.

In the optimization approach, \cite{Wheatland+etal+2000} defined a quantity L:	
\begin{equation}
L = \int _V[B^{-2}|(\nabla \times \vec{B}) \times \vec{B}|^2 + |\nabla \cdot \vec{B}|^2]dV ,
\label{eq:LebsequeI}
\end{equation}

 where $\vec{B}$ is defined in a volume V. If L is decreased to zero, equations (1)-(3) are fulfilled, then the field is force-free in the volume V. In order to reduce L, $\vec{B}$ needs to be evolved like:
\begin{equation}
\frac{\partial \vec{B}}{\partial t} = \mu \vec{F}.
\label{eq:LebsequeI}
\end{equation}
\cite{Wheatland+etal+2000} provided a testing to optimization method. \cite{Inhester+Wiegelmann+2006} provided detailed comparison of optimization method and Grad-Rubin method by implementing these two algorithms and comparing the performance. In addition, \cite{Liu+etal++2011+3} used two semi-analytical solutions of force-free fields to test two other NLFF extrapolation methods: boundary integral equation (BIE) method developed by \cite{Yan+Sakurai+2000} and approximate vertical integration (AVI) method developed by \cite{Song+etal+2006}.

\cite{Wiegelmann+2004} improved the optimization approach by showing how the magnetic field can be reconstructed only from the button boundary and developed a code which will be used later.

Moreover, although the NLFF field model is widely-used (\citealt{R¨¦gnier+Amari+2004}, \citealt{Wiegelmann+etal+2005}, \citealt{Schrijver+2008}), a joint study of \cite{DeRosa+etal+2009} concluded that a successful application of NLFF field extrapolation should satisfy several requirements.  Recently \cite{Wiegelmann+etal+2012} offered a detailed discussion on this problem and proved that their results fulfill these requirements. The results of nonlinear force-free modeling should be used with some caution.

\section{Calculating Induced Electric Field Distribution}

In solar atmosphere, the static electric field is neglected due to the plasma screening. We consider the case that there is no accumulation of charged particles in the solar atmosphere, and the electric field is mainly generated from the time variation of magnetic field. That is,
\begin{equation}
\nabla \cdot \vec{E} = 0 ,
\label{eq:LebsequeI}
\end{equation}
and
\begin{equation}
\nabla \times \vec{E} = \Vec{\Omega}
\label{eq:LebsequeI}
\end{equation}
where
\begin{equation}
\Vec{\Omega} = -\frac{\partial \vec{B}}{\partial t} .
\label{eq:LebsequeI}
\end{equation}

We could derive electric field directly from the time variation of magnetic field (\citealt{Batchelor+2000}):
\begin{equation}
\vec{E} = \frac{1}{4\pi}\iiint _D \frac{\Vec{\Omega}(\xi,\eta,\zeta) \times \vec{R}}{R^3}d\xi d\eta d\zeta.
\label{eq:LebsequeI}
\end{equation}

In order to get $\vec{E}$ from equation (9), we introduce the Monte Carlo method which is a statistical simulation method to solve the triple integral problem. This method can be used to approximate the involved integral if the precise value of integrals is not important and its estimated value is enough, or if the precise value is unable to get.

If f(x,y,z) is continuous function on domain D, and g(x,y,z) is a probability density function such that
\begin{equation}
\iiint _D g(x,y,z) dxdydz= 1,
\label{eq:LebsequeI}
\end{equation}
$(x_i,y_i,z_i)(i=1,2,...N)$ are a sequence of random numbers that fall in domain D, according to the theorem of Monte Carlo method, when N is large enough, we have
\begin{equation}
\iiint _D f(x,y,z) dxdydz \approx \frac{1}{N} \sum _{i=1} ^N \frac{f(x_i,y_i,z_i)}{g(x_i,y_i,z_i}.
\label{eq:LebsequeI}
\end{equation}

If g(x,y,z) is constant, equation (10) becomes
\begin{equation}
g(x,y,z)\iiint _D dxdydz= \frac{1}{D},
\label{eq:LebsequeI}
\end{equation}
and
\begin{equation}
\iiint _D f(x,y,z) dxdydz \approx \frac{D}{N} \sum _{i=1} ^N f(x_i,y_i,z_i).
\label{eq:LebsequeI}
\end{equation}.
By applying Monte Carlo method to equation (9), we get the final equation of $\vec{E}$ as
\begin{equation}
\vec{E}\approx \frac{D}{N \cdot 4\pi}\sum _{i=1} ^N\frac{\Vec{\Omega}(\xi_i,\eta_i,\zeta_i) \times \vec{R}}{R^3}
\label{eq:LebsequeI}
\end{equation}.

\section{The Induced Electric Field in NOAA AR11072}

To implement the method we described in the previous sections, we provide an example here. At first we extrapolate magnetic field of photosphere from magnetograms, then we compute the time different of two magnetic field distributions $\Vec{\Omega} = -\partial \vec{B} / \partial t$, at last we use Monte Carlo method to calculate the electric field distribution.

\begin{figure}[!t]
   \centering
  \includegraphics[width=4cm, height=4cm, angle=0]{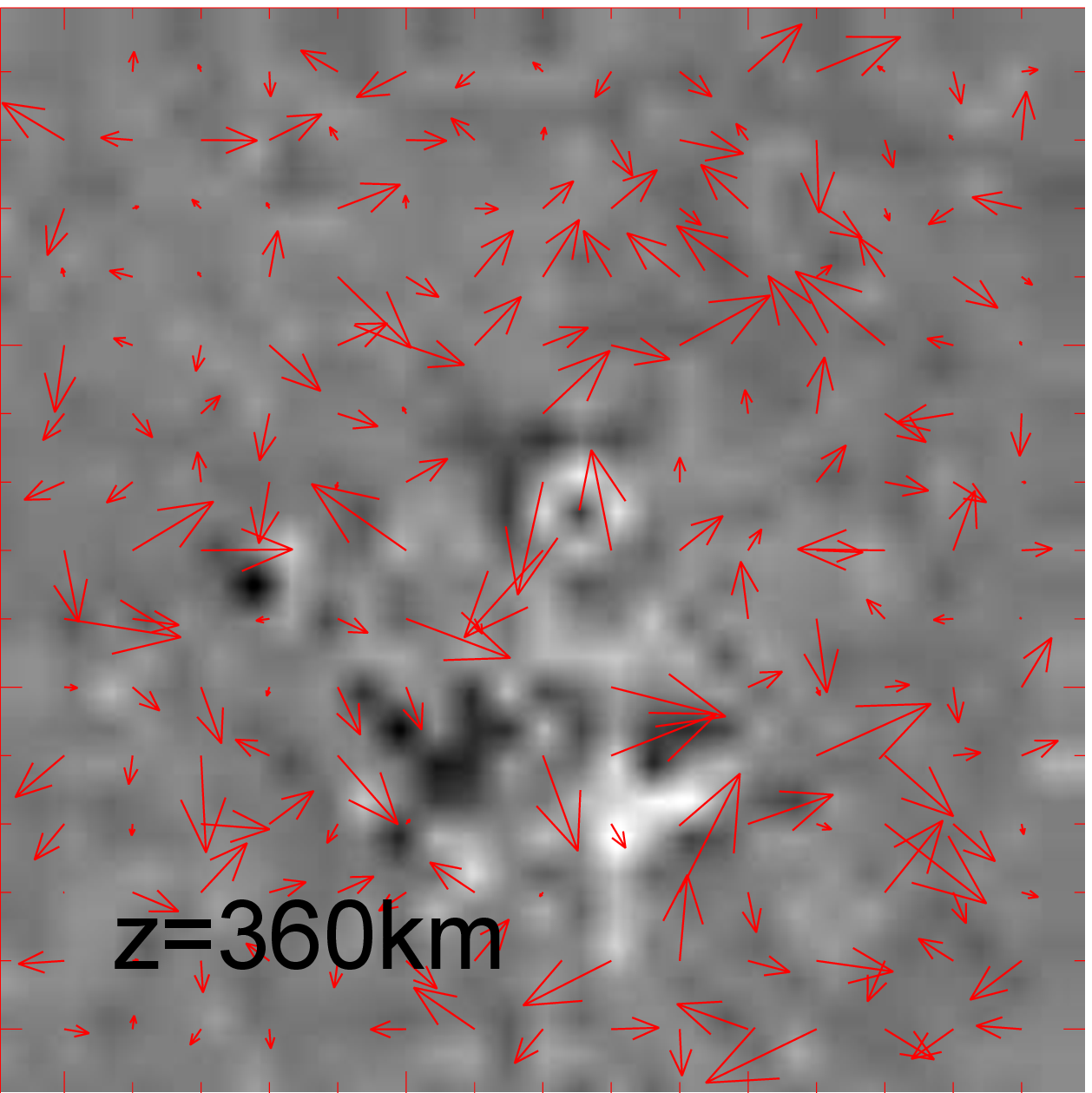}
  \includegraphics[width=4cm, height=4cm, angle=0]{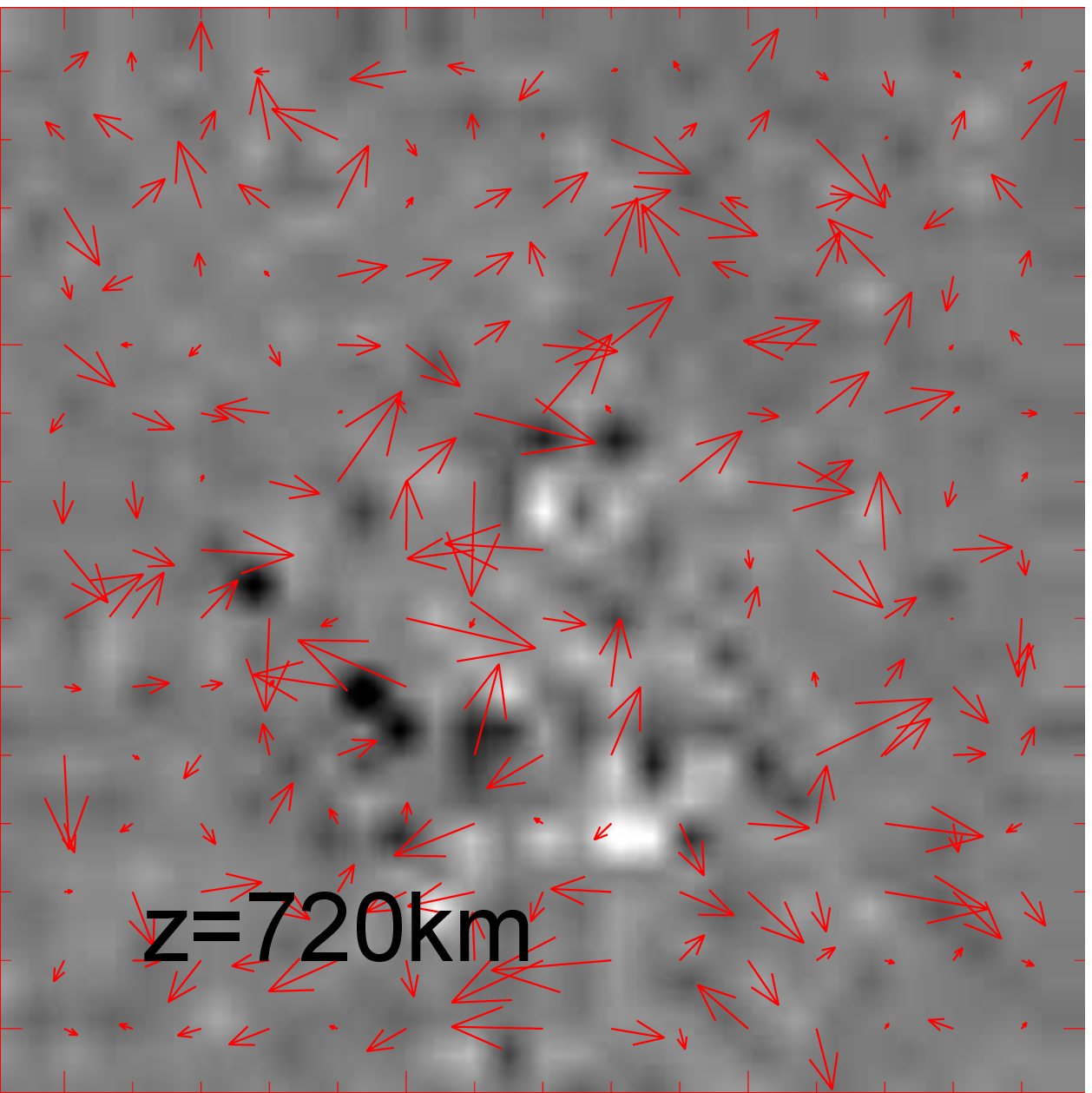}
  \includegraphics[width=4cm, height=4cm, angle=0]{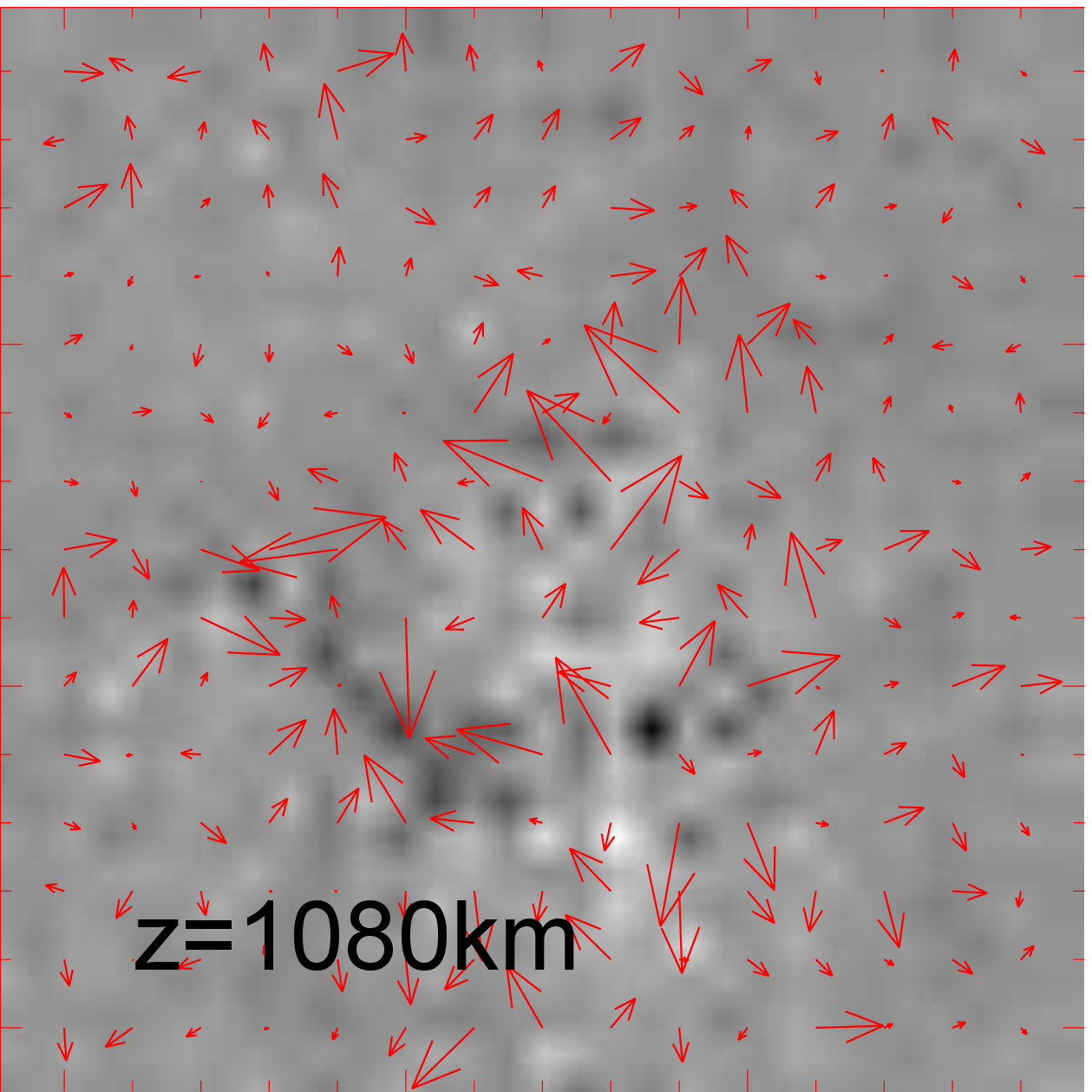}
  \includegraphics[width=4cm, height=4cm, angle=0]{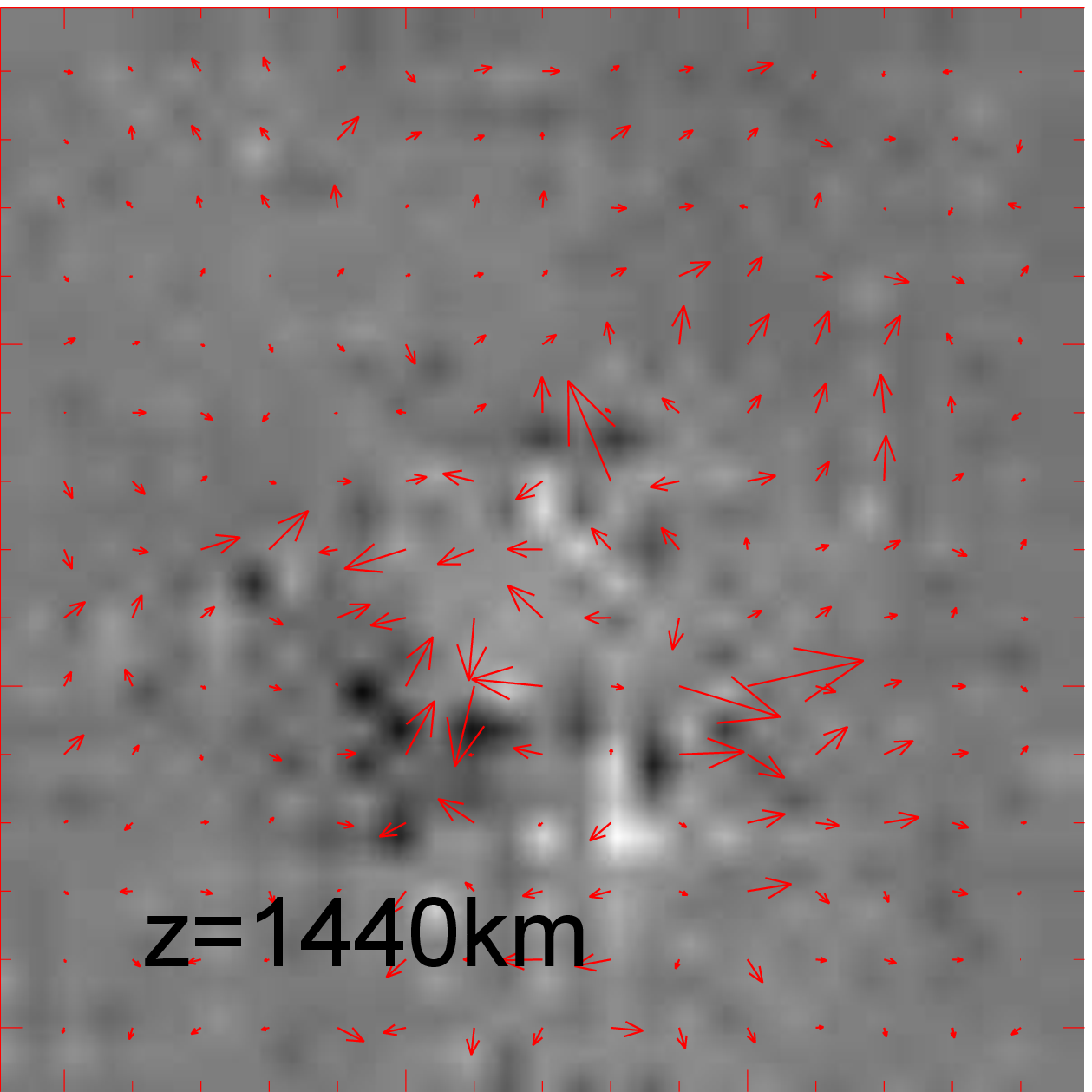}
  \includegraphics[width=4cm, height=4cm, angle=0]{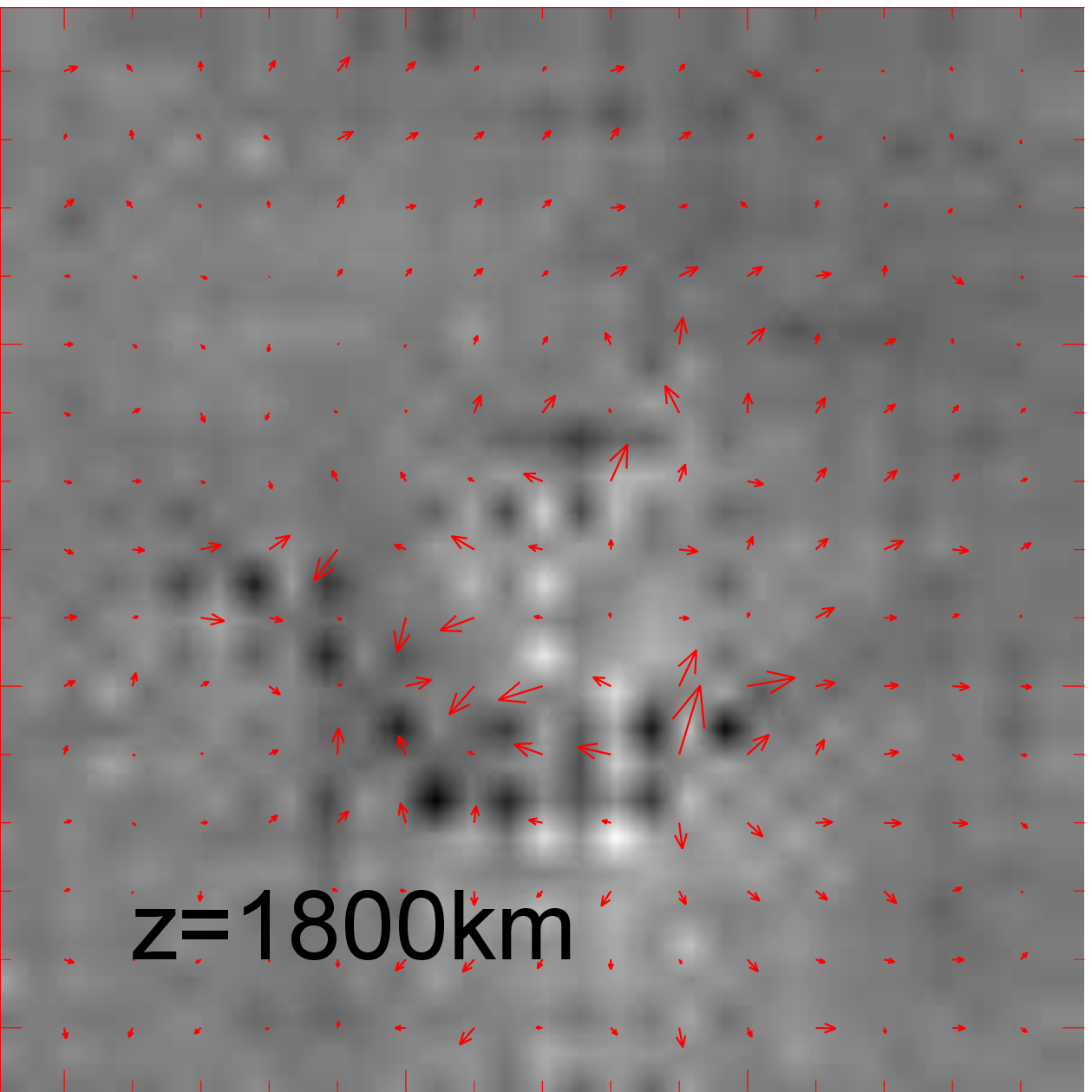}
  \includegraphics[width=4cm, height=4cm, angle=0]{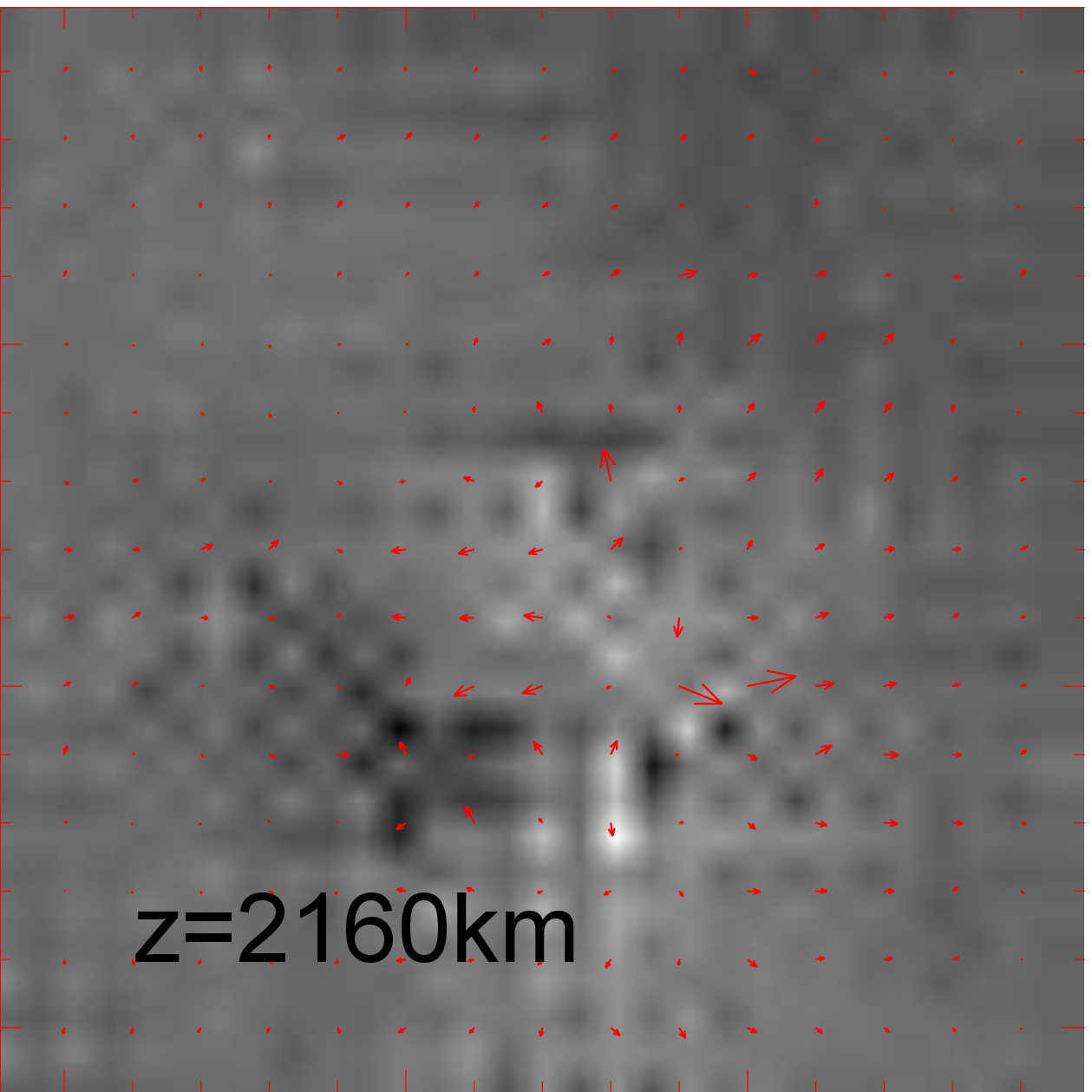}
   \caption{Extrapolated results of a magnetogram for Active Region 11072 observed from HMI/SDO at 12:12 on May 20th, 2010. The size of every image is 10.8 $\times$ 10.8 $Mm^2$, and the value of z means the distance from photosphere in the unit of km. Arrows show directions and amplitudes of $B_x$ and $B_y$, while the background image shows the amplitude of $B_z$}
   \label{Fig1}
   \end{figure}

\begin{table}[b]
\bc
\begin{minipage}[]{100mm}
\caption[]{Maximum and average value of $\vec{B}$ at different layers
\label{mbh}}\end{minipage}
\setlength{\tabcolsep}{2.5pt}
\small
 \begin{tabular}{ccccccccccccc}
  \hline\noalign{\smallskip}
Z &  $max(|B_x|)$ & $max(|B_y|)$ & $max(|B_z|)$ &  $mean(B_x)$ & $mean(B_y)$ & $mean(B_z)$ \\
(km) & & & (G) & & &\\
  \hline\noalign{\smallskip}
360 & $1\times 10^3$ & $8.4\times 10^2$ & $8.3\times 10^2$ & $4\times 10^{-1}$ & $-1.1\times 10^{-1}$ & -3.2\\
720 & $1.1\times 10^3$ & $7.8\times 10^2$ & $1.1\times 10^3$ & $7.6\times 10^{-1}$& $-1.2\times 10^{-1}$ & -3.4\\
1080 & $7.2\times 10^2$ & $1.1\times 10^3$ & $2.7\times 10^3$ & $6.4\times 10^{-1}$ & $-9.6\times 10^{-2}$ & -3.2\\
1440 & $8.2\times 10^2$ & $1\times 10^3$ & $4.5\times 10^2$ & $4.5\times 10^{-1}$ & $-3.3\times 10^{-2}$ & -3.3\\
1800 & $1.9\times 10^2$ & $4.2\times 10^2$ & $8.6\times 10^2$ & $3.9\times 10^{-1}$ & $-4.4\times 10^{-2}$ & -3\\
2160 & $2.7\times 10^2$ & $2.7\times 10^2$ & $8.1\times 10$ & $3.2\times 10^{-1}$& $-3.3\times 10^{-2}$ & -3 \\
  \noalign{\smallskip}\hline
\end{tabular}
\ec
\end{table}

In this example, we use observational data from HMI/SDO which provides high spatial and temporal resolution vector magnetograms(\citealt{Schou+etal+2012}). HMI provides continuous vector magnetograms at a 12-minute cadence and released in the past two years several data series of cutouts of the original full disk images. The full released data summary can be found at the SDO Joint Science Operations Center web page (http://jsoc.stanford.edu). From the available data releases, we chose the 6-day cutouts of 512$\times$512 pixels for NOAA Active Region 11072 from May 20th 2010 to May 26th 2010.

We run the code developed by \cite{Wiegelmann+2004} (optimization method) to extrapolate magnetic field from vector magnetograms, and both potential and NLFF field are generated, \cite{Liu+etal++2011+5} provided detailed comparison of NLFF Field and Potential Field. We extrapolate magnetic field to a volume of 512 $\times$ 512 $\times$ 9 pixels and the distance between two adjacent pixels is 0.5 arc seconds. Figure 1 shows the extrapolated NLFF field of one magnetogram.

The horizontal cuts of the lower six layers consisting 30 $\times$ 30 pixels are shown in Figure 1. The value of z indicates height from the photosphere in unit of kilometer, arrows show directions and amplitudes of $B_x$ and $B_y$, while the background image shows the amplitude of $B_z$.

Table 1 shows the maximum and average value of the three components of the extrapolated magnetic field, and the first image of Figure 5 shows the average of absolute value of magnetic field at six layers. We show in Figure 2 the difference between two extrapolated results, that is, $\vec{B}_2$ - $\vec{B}_1$ where $\vec{B}_1$ is the magnetic field in Figure 1.

\begin{figure}[h]
   \centering
  \includegraphics[width=4cm, height=4cm, angle=0]{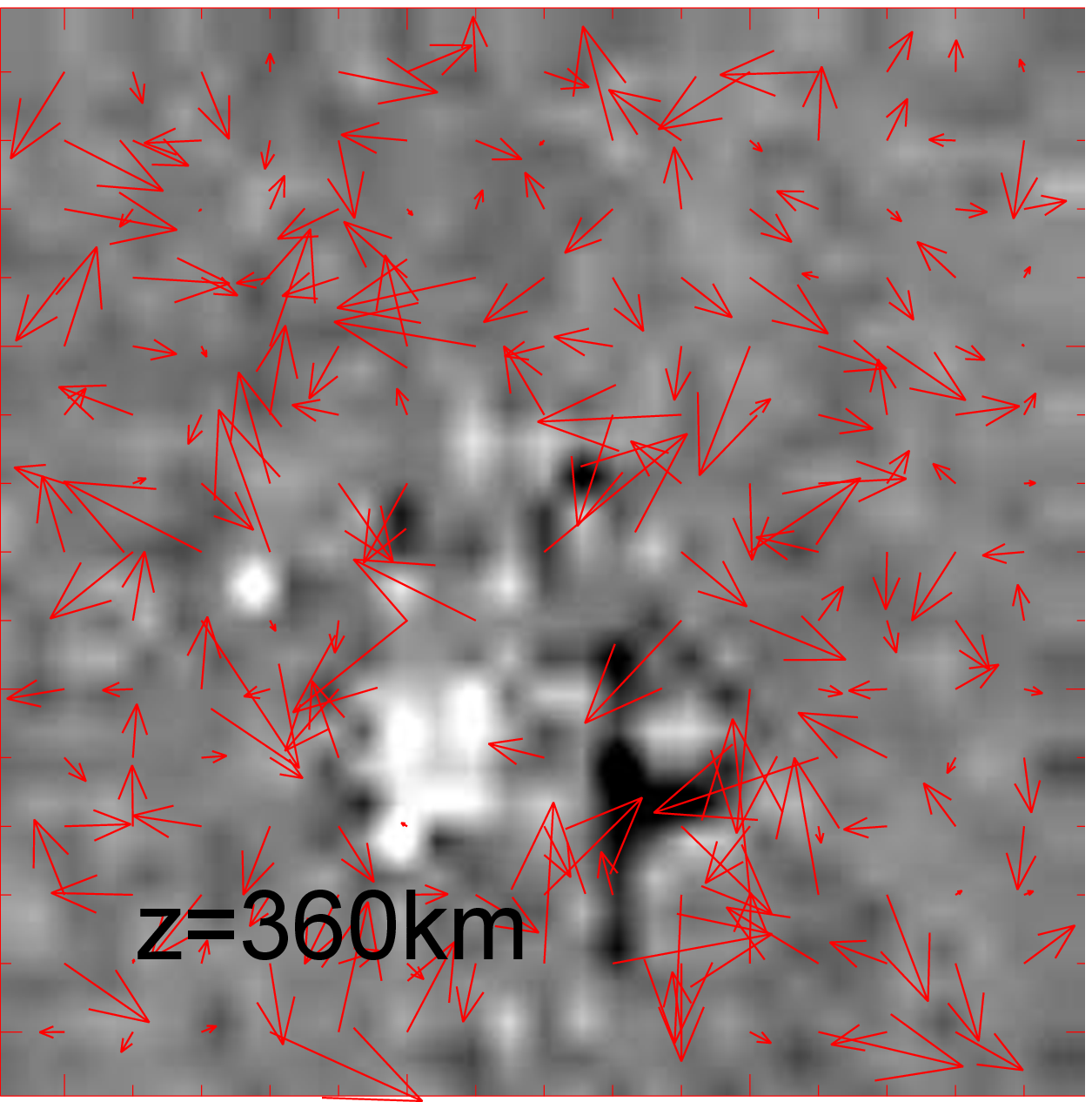}
  \includegraphics[width=4cm, height=4cm, angle=0]{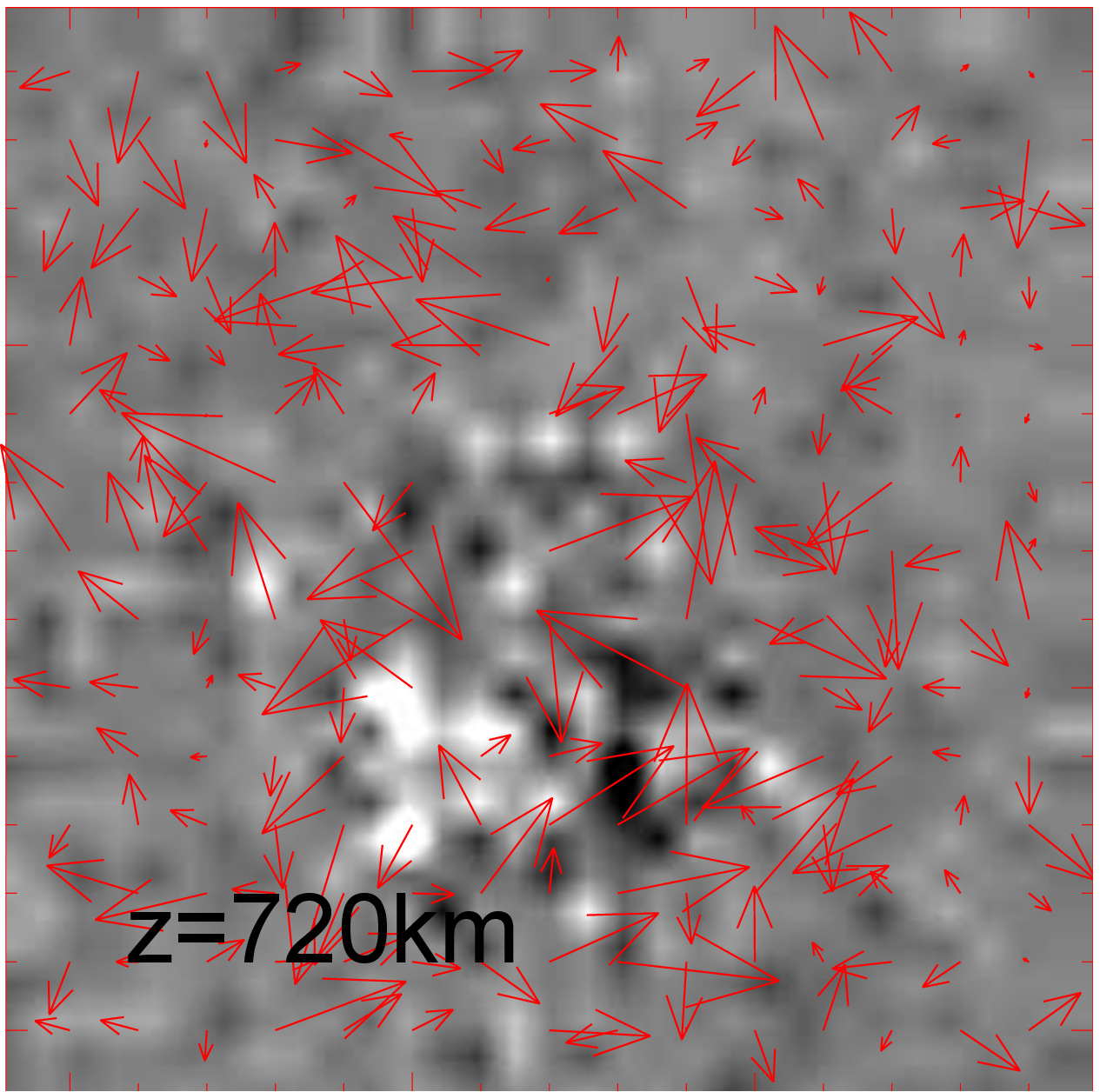}
  \includegraphics[width=4cm, height=4cm, angle=0]{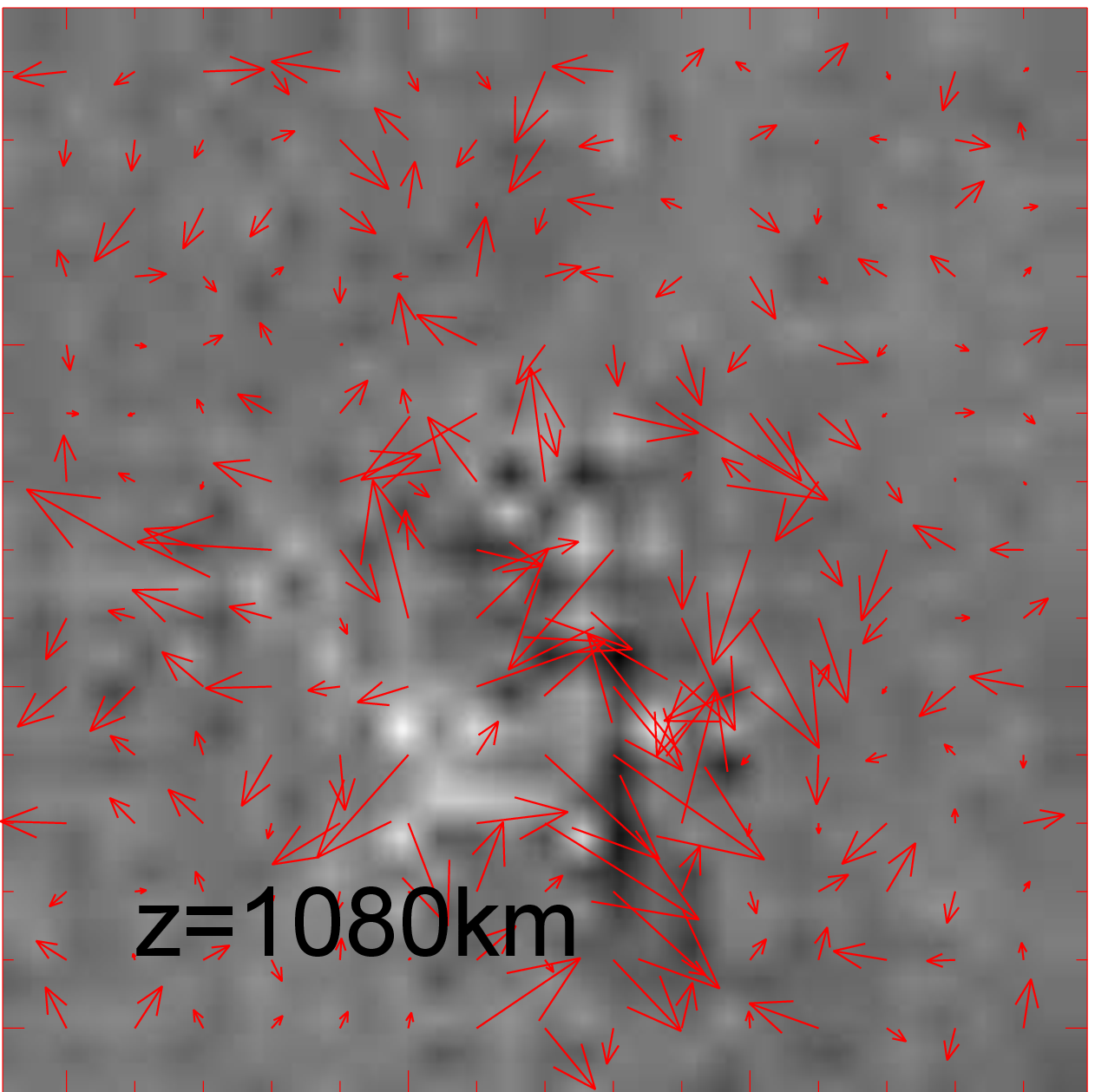}
  \includegraphics[width=4cm, height=4cm, angle=0]{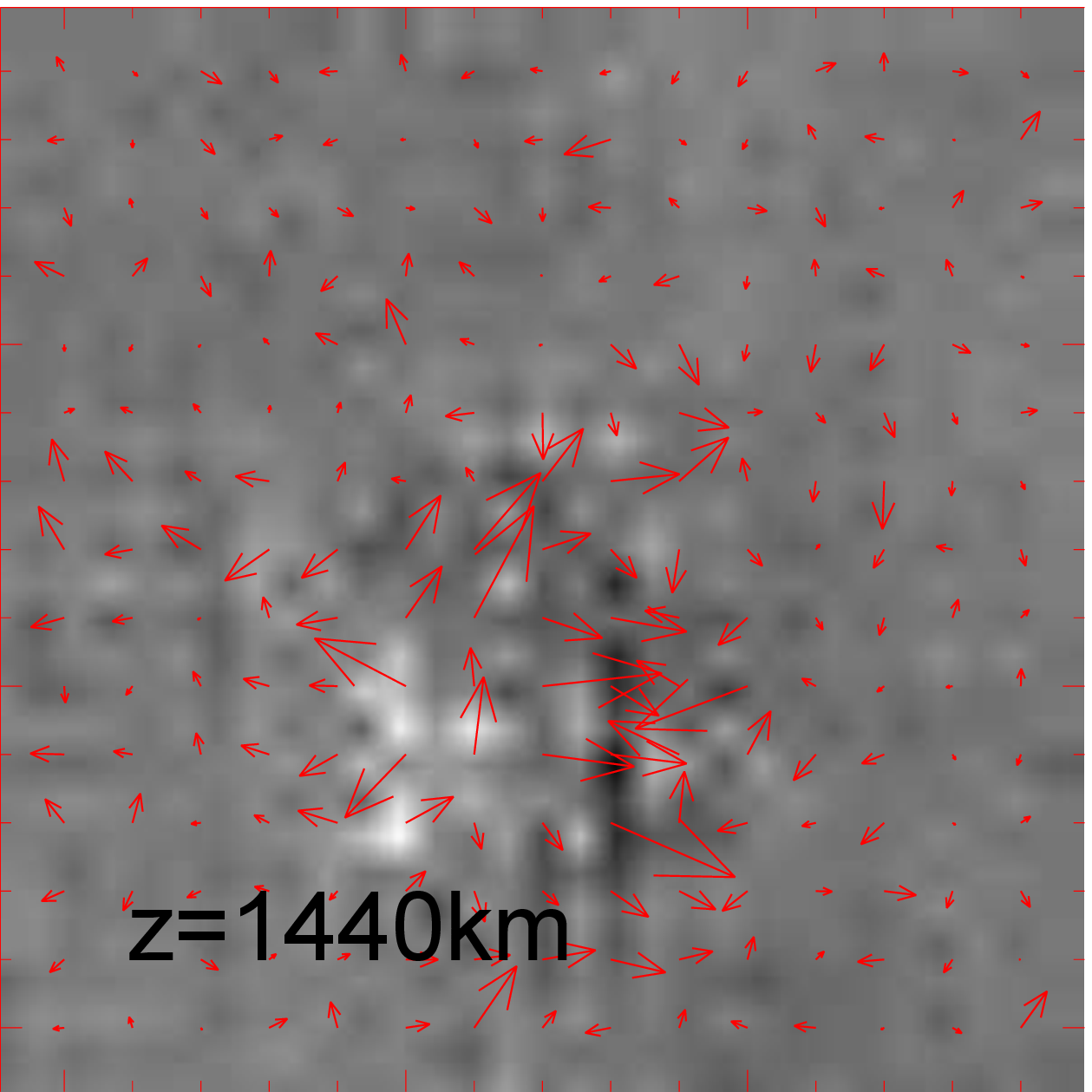}
  \includegraphics[width=4cm, height=4cm, angle=0]{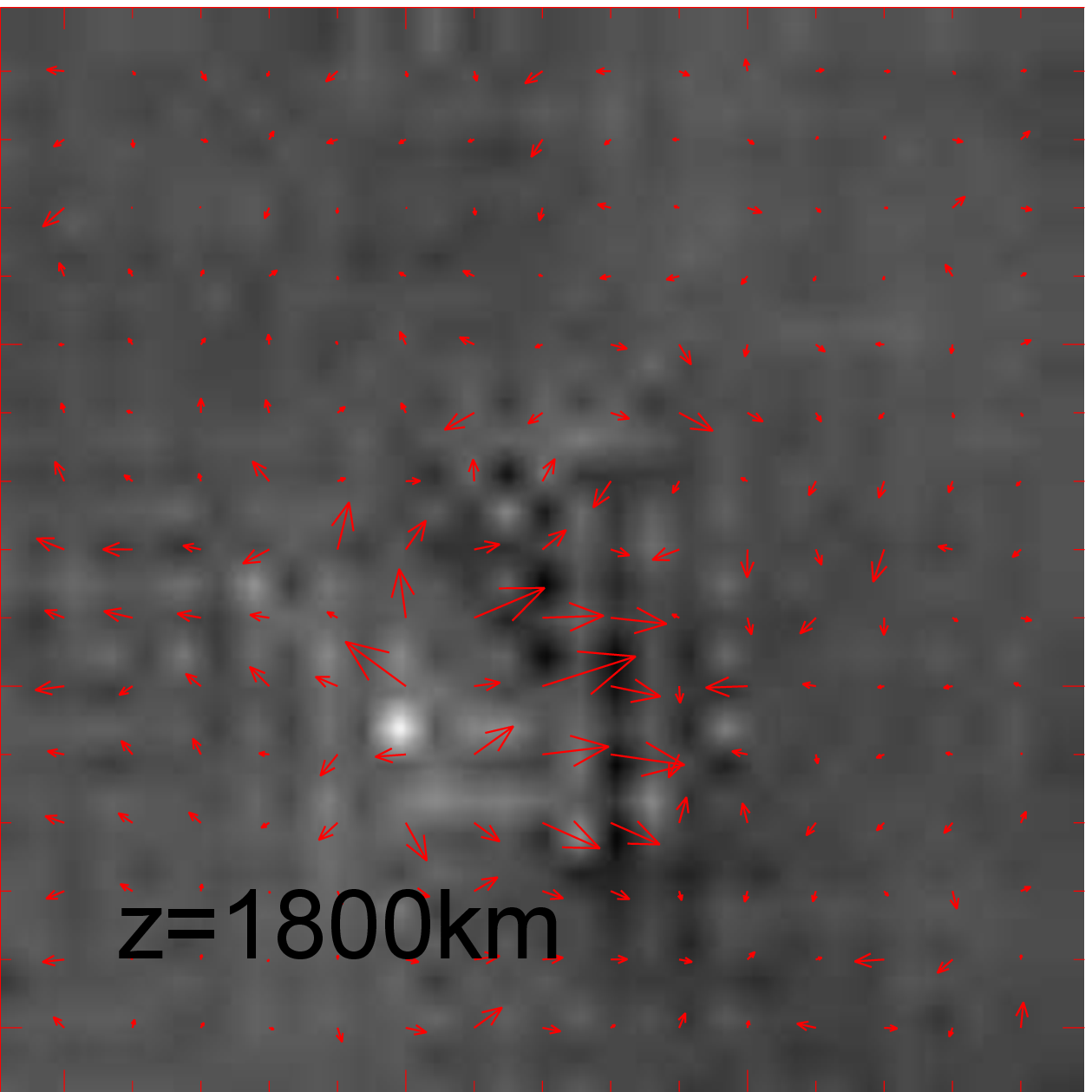}
  \includegraphics[width=4cm, height=4cm, angle=0]{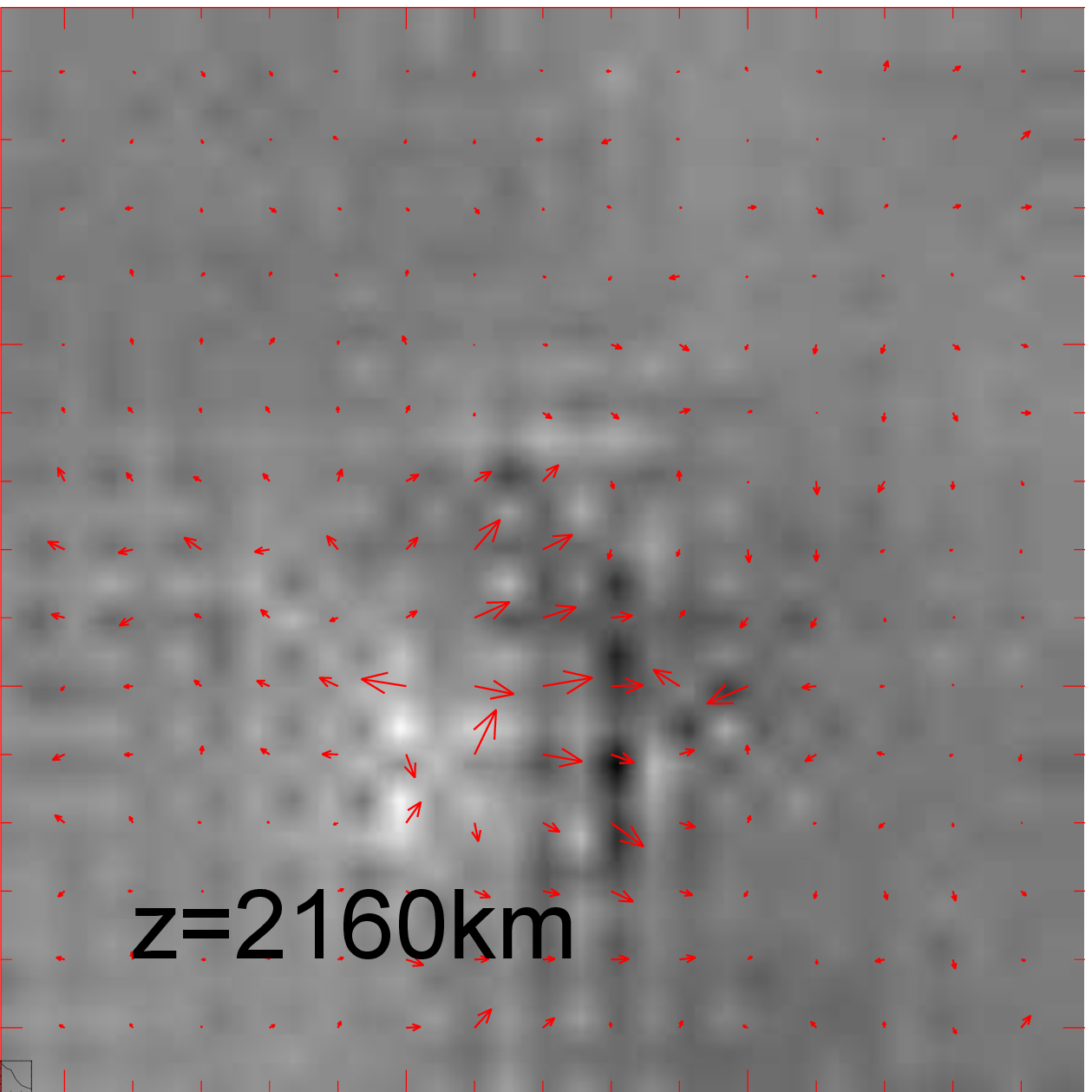}
   \caption{Time variation of magnetic field extrapolated from two continuous vector magnetograms, the first one is observed at 12:12 on May 20th, 2010 by HMI/SDO, and the other is observed 12 minutes later. }
   \label{Fig2}
   \end{figure}

\begin{table}[h]
\bc
\begin{minipage}[]{100mm}
\caption[]{Maximum and average value of $\Delta \vec{B}$ at different layers
\label{mbh}}\end{minipage}
\setlength{\tabcolsep}{2.5pt}
\small
 \begin{tabular}{ccccccccccccc}
  \hline\noalign{\smallskip}
Z &  $max(|\Delta B_x|)$ & $max(|\Delta B_y|)$ & $max(|\Delta B_z|)$ &  $mean(\Delta B_x)$ & $mean(\Delta B_y)$ & $mean(\Delta B_z)$ \\
 (km)& & & (G) & & &\\
  \hline\noalign{\smallskip}
360 & $1.2\times 10^3$ & $1.4\times 10^3$ & $1.2\times 10^3$ & $1.5\times 10^{-1}$ & $1.7\times 10^{-1}$ & $4.3\times 10^{-2}$ \\

720 & $1.1\times 10^3$ & $1\times 10^3$ & $1.3\times 10^3$ & $2.2\times 10^{-1}$ & $2.7\times 10^{-3}$ & $4.1\times 10^{-2}$ \\

1080 & $7.4\times 10^2$ & $1.1\times 10^3$ & $2.7\times 10^3$ & $-1.7\times 10^{-1}$ & $1.6\times 10^{-1}$ & $4.2\times 10^{-2}$ \\

1440 & $8.1\times 10^2$ & $1\times 10^3$& $4.8\times 10^2$ & $-8.9\times 10^{-2}$ & $-6.4\times 10^{-3}$ & $3.9\times 10^{-2}$ \\

1800 & $1.9\times 10^2$ & $4.2\times 10^2$ & $8.7\times 10^2$ & $-8.8\times 10^{-2}$ & $6.8\times 10^{-2}$ & $3.7\times 10^{-2}$ \\

2160 & $2.7\times 10^2$ & $2.6\times 10^2$ & $8.9\times 10$ & $-9\times 10^{-2}$ & $3.7\times 10^{-2}$ & $3.6\times 10^{-2}$ \\
  \noalign{\smallskip}\hline
\end{tabular}
\ec
\end{table}

Table 2 shows the maximum and average value of the difference of the extrapolated magnetic field, and the second image of Figure 5 shows the average of absolute value at six layers. Monte Carlo method requires that $\Omega=-\frac{\partial \vec{B}}{\partial t}$ in equation (9) and (14) is spatial continuous function, but our observed and extrapolated $\Omega$ is discrete with spatial distance of 0.5 arc seconds. Thus, we linearize $\Omega$ in unit of 0.5 arc seconds to reconstruct a continuous function.

\begin{figure}[h]
   \centering
  \includegraphics[width=4.5cm, angle=0]{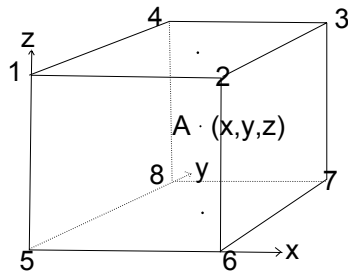}
   \caption{The $\Omega$ at the point A(x,y,z) is linearized from 8 points next to it. The position of these adjacent points is ($x_i,y_i,z_i$, i = 1...8), and every edge of the cube represents 0.5 arc seconds. A has two projections in the upper and lower surfaces where the $\Omega$ should be computed at first.}
   \label{Fig3}
   \end{figure}

We use the linear algorithm below to compute $\Omega(x,y,z_up)$ of the upper projection in Figure 3:
\begin{eqnarray*}
\Omega(x,y_{back},z_{up}) = \Omega(x_{left},y_{back},z_{up}) \times (x_{right}-x)/0.5) + \Omega(x_{right},y_{back},z_up) \times (x-x_{left})/0.5,\\
\Omega(x,y_{forth},z_{up}) = \Omega(x_{left},y_{forth},z_{up}) \times (x_{right}-x)/0.5) + \Omega(x_{right},y_{forth},z_{up}) \times (x-x_{left})/0.5,\\
\Omega(x,y,z_{up}) = \Omega(x,y_{back},z_{up}) \times (y_{forth}-y)/0.5 +\Omega(x,y_{forth},z_{up}) \times (y-y_{back})/0.5,
\end{eqnarray*}
where
\begin{eqnarray*}
x_{left} = x_1 = x_4 = x_5 = x_8, x_{right} = x_2 = x_3 = x_6 = x_7,\\
y_{back} = y_1 = y_2 = y_5 = y_6, y_(forth) = y_3 = y_4 = y_7 = y_8,\\
z_{down} = z_5 = z_6 = z_7 = z_8, z_{up} = z_1 = z_2 = z_3 = z_4.
\end{eqnarray*}

\begin{figure}[h]
   \centering
  \includegraphics[width=4cm, height=4cm, angle=0]{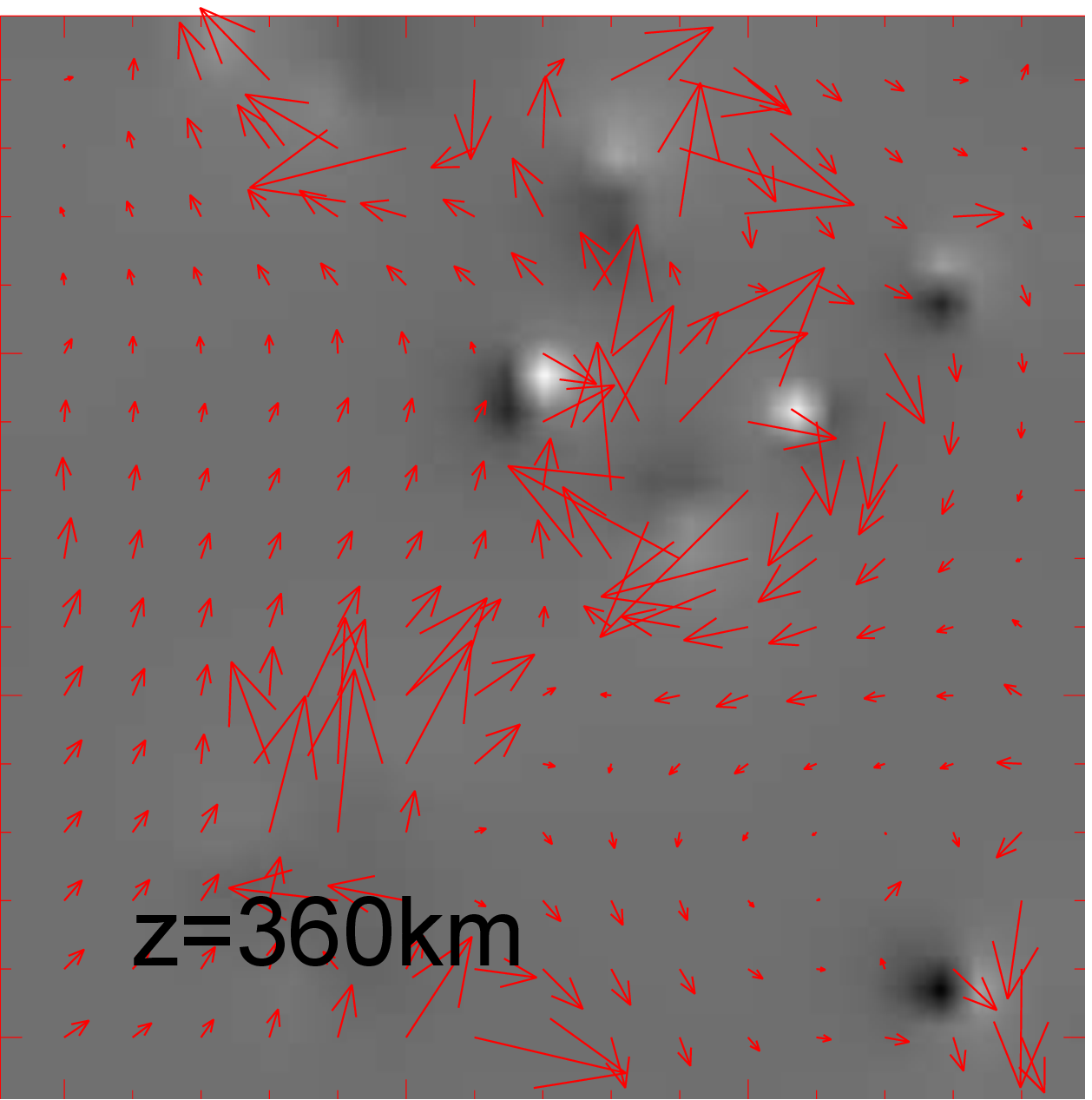}
  \includegraphics[width=4cm, height=4cm, angle=0]{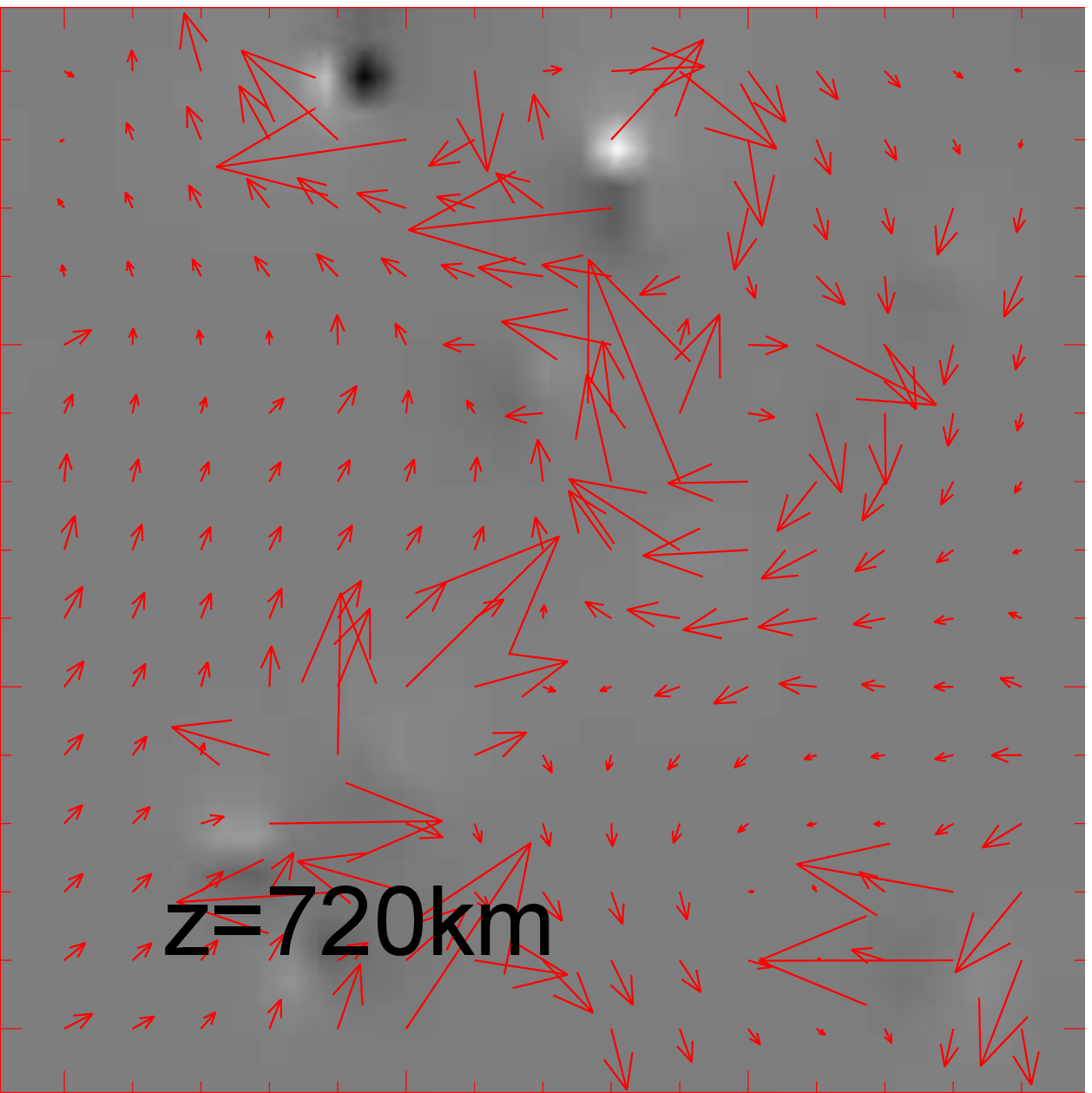}
  \includegraphics[width=4cm, height=4cm, angle=0]{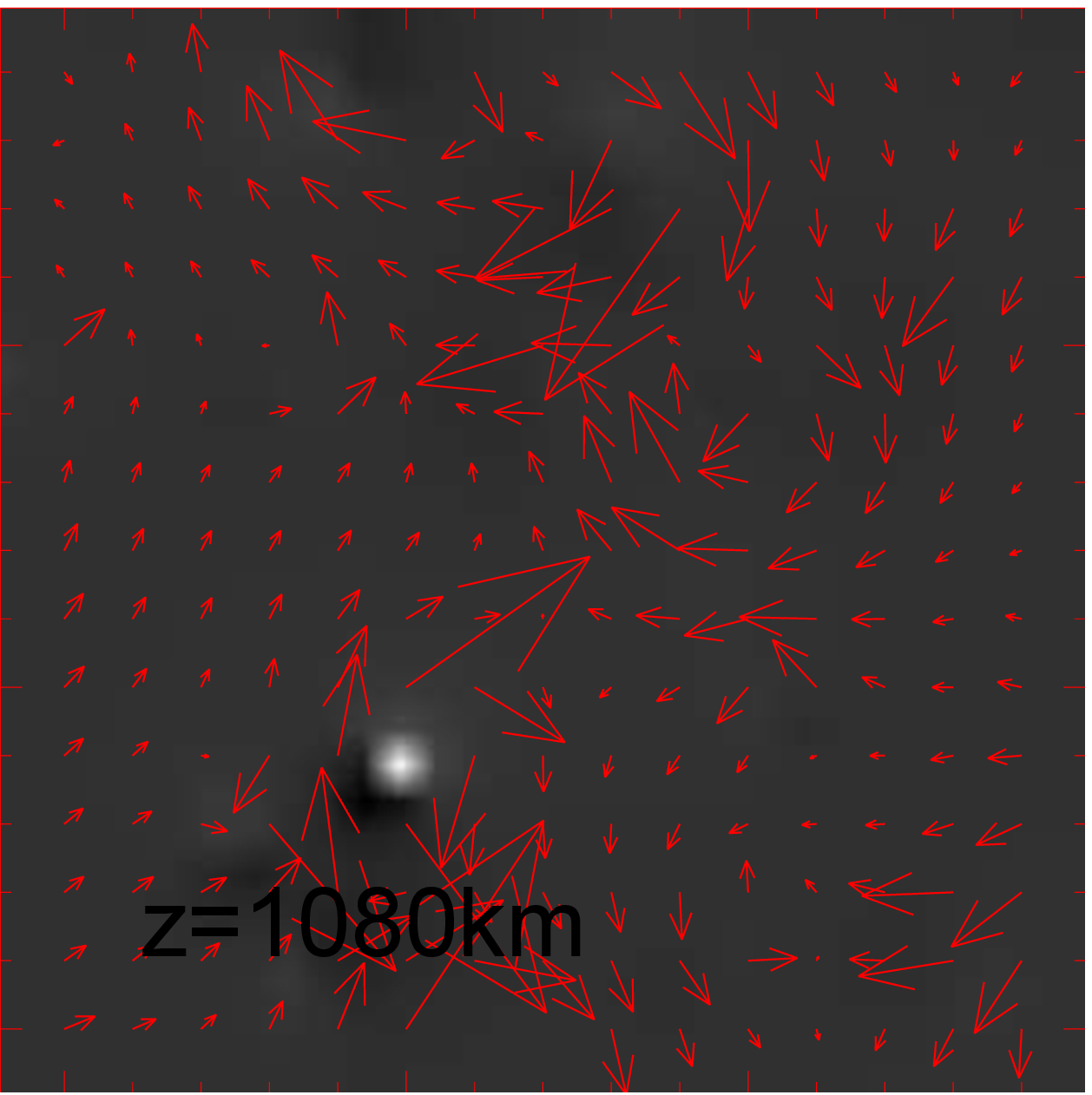}
  \includegraphics[width=4cm, height=4cm, angle=0]{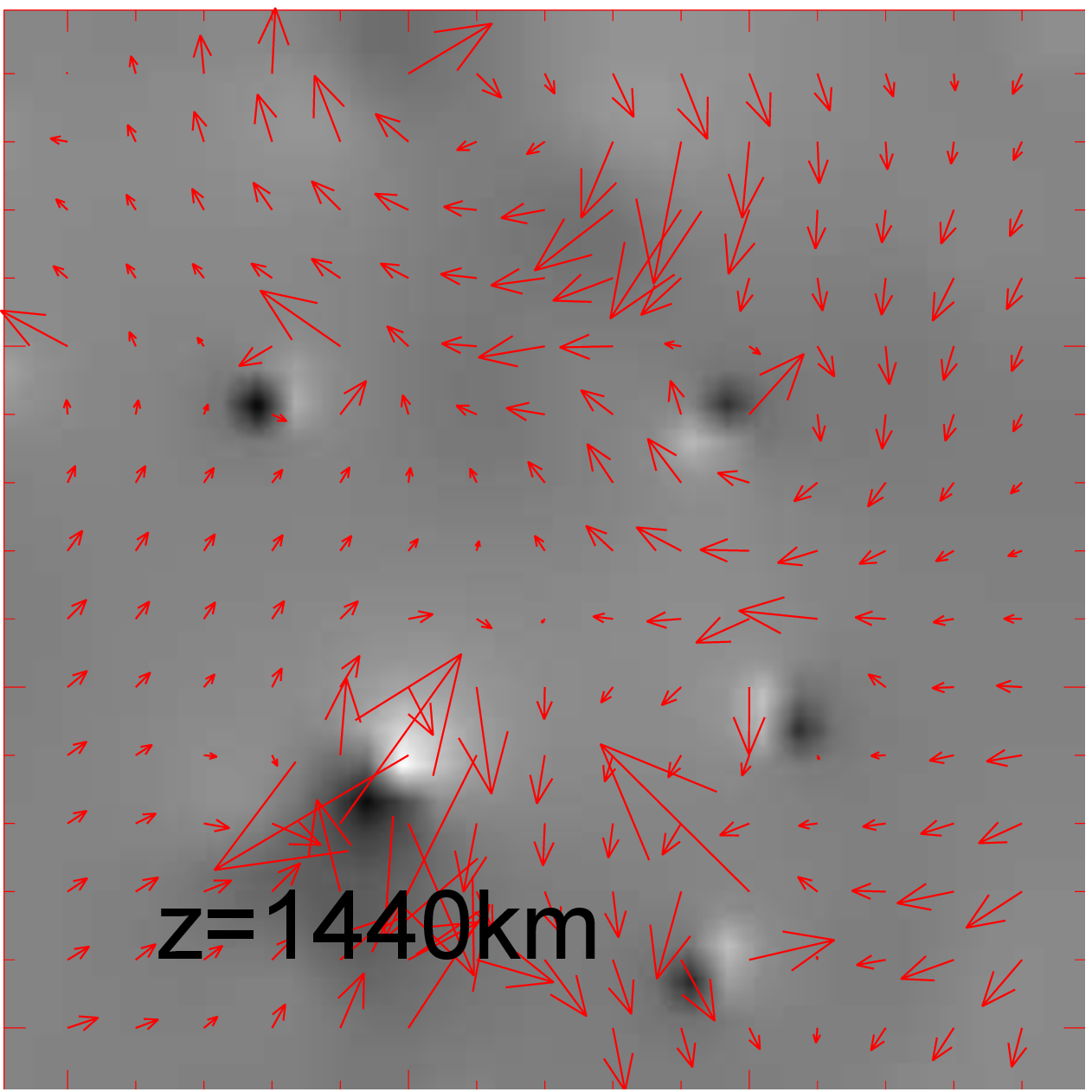}
  \includegraphics[width=4cm, height=4cm, angle=0]{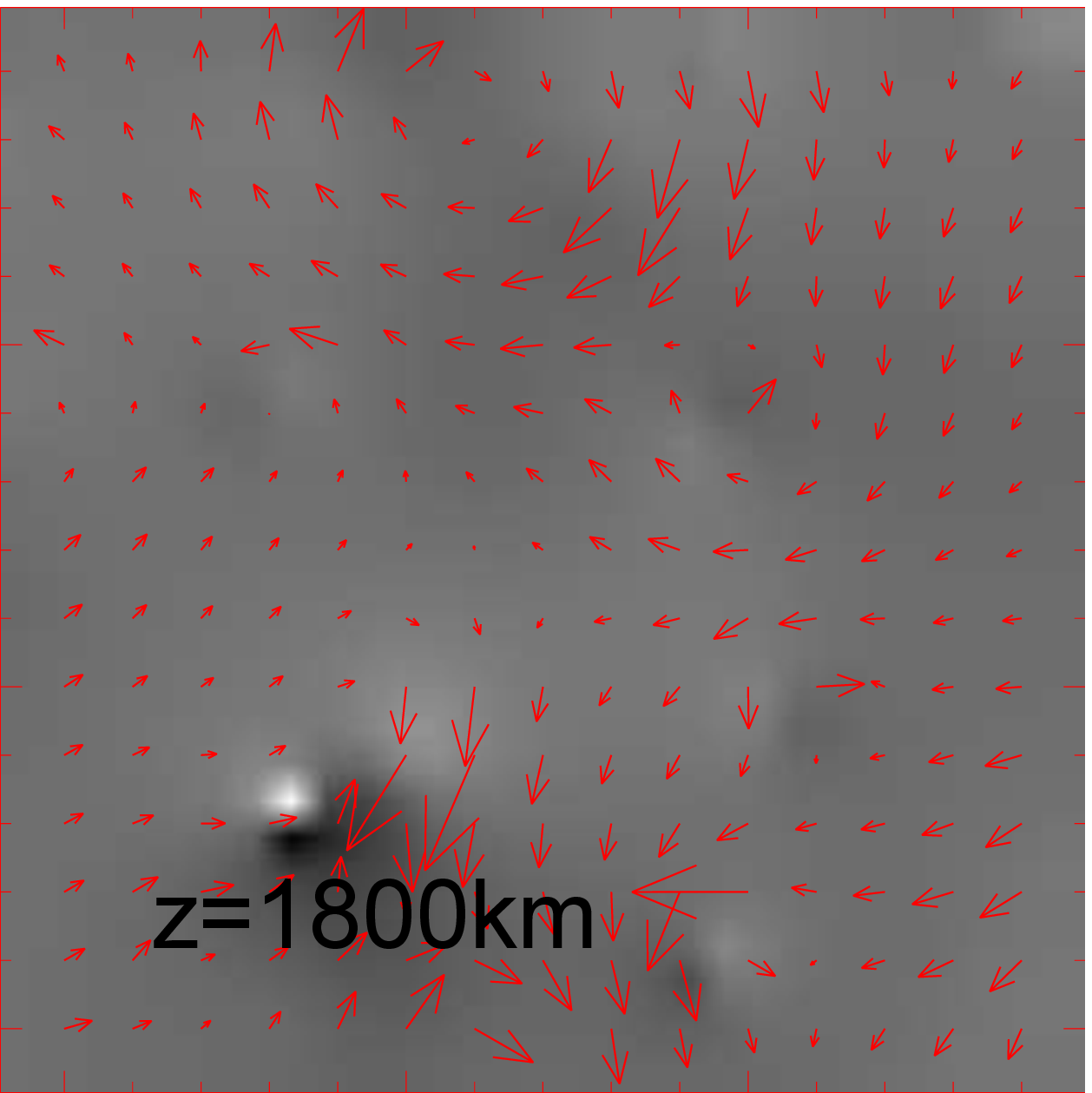}
  \includegraphics[width=4cm, height=4cm, angle=0]{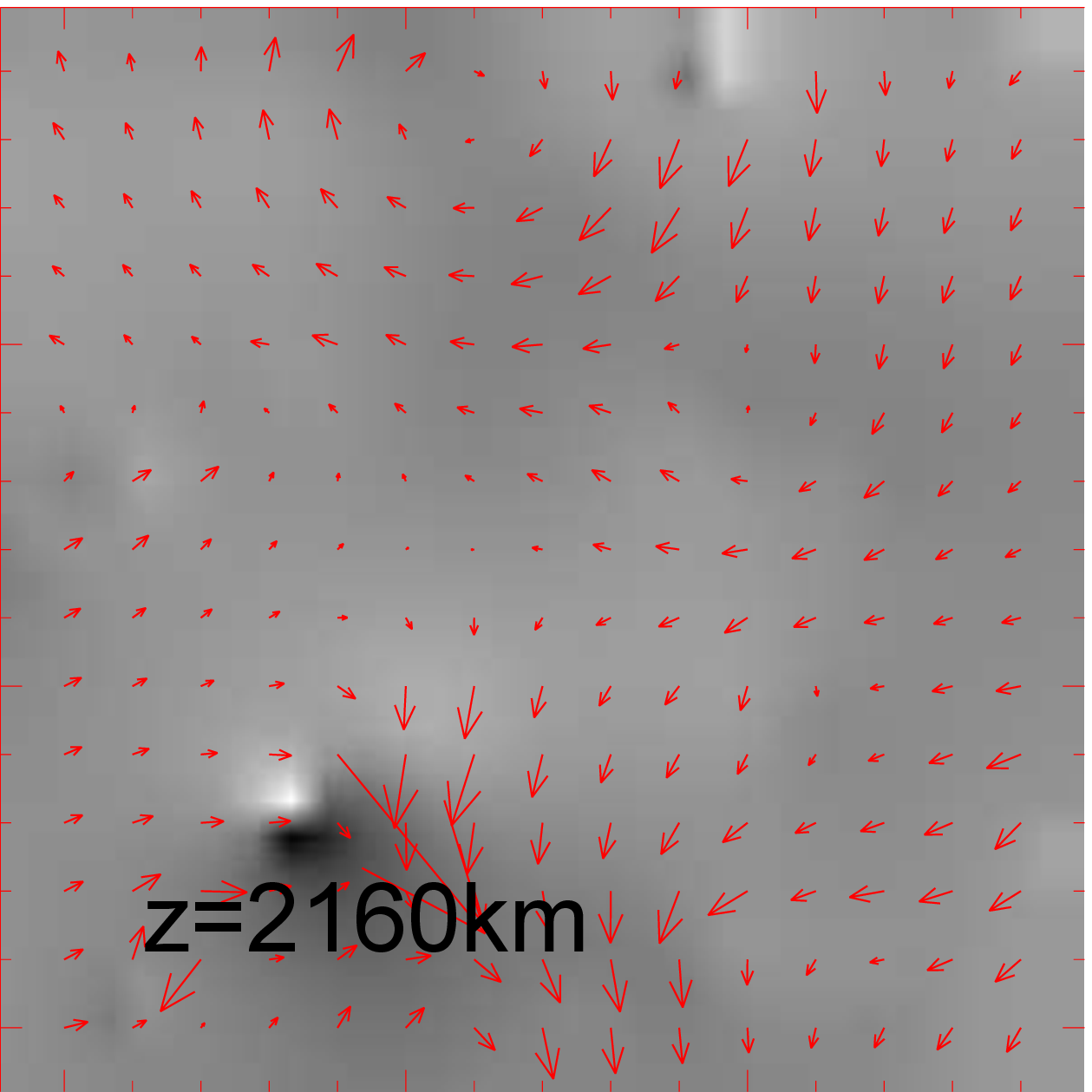}
   \caption{Computed ``vector electrograms'' (10.8 $\times$ 10.8 $Mm^2$) of different depth. Arrows show directions and amplitudes of $E_x$ and $E_y$, while the background image shows the amplitude of $E_z$}
   \label{Fig4}
   \end{figure}

Similarly, we can get $\Omega(x,y,z_{down})$, together with $\Omega(x,y,z_{up})$, and we compute the linearized $\Omega(x,y,z)$ from
\begin{eqnarray*}
\Omega(x,y,z) = \Omega(x,y,z_{down}) \times (z_{up} - z)/0.5 + \Omega(x,y,z_{up}) \times (z - z_{down})/0.5.
\end{eqnarray*}
For saving time, we only generate 10,000 random points in the cube of 512 $\times$ 512 $\times$ 7, and calculate the distribution of induced electric field depending on the equation (14). One of our result are showed in Figure 4.

\begin{table}[h]
\bc
\begin{minipage}[]{100mm}
\caption[]{Maximum and average value of $\vec{E}$ at different layers
\label{mbh}}\end{minipage}
\setlength{\tabcolsep}{2.5pt}
\small
 \begin{tabular}{ccccccccccccc}
  \hline\noalign{\smallskip}
Z &  max($|E_x|$) & max($|E_y|$) & max($|E_z|$) &  mean($E_x$) & mean($E_y$) & mean($E_z$) \\
 (km)& & & (V/cm) & & &\\
  \hline\noalign{\smallskip}
 360 & $5.5\times 10^{2}$ & $3.9\times 10^{2}$ & $2.5\times 10^{2}$ & $3.5\times 10^{-3}$ & $1.6\times 10^{-3}$ & $1.1\times 10^{-3}$ \\

 720 & $3.6\times 10^{2}$ & $1.7\times 10^{2}$ & $4.4\times 10^{2}$ & $9\times 10^{-4}$ & $2.3\times 10^{-4}$ & $-1.4\times 10^{-3}$ \\

 1080 & $4.6\times 10^{2}$ & $2.3\times 10^{2}$ & $1.6\times 10^{2}$ & $2.1\times 10^{-3}$ & $1.2\times 10^{-3}$ & $7.9\times 10^{-4}$ \\

 1440 & $8.1\times 10$ & $2\times 10^{2}$ & $1.1\times 10^{2}$ & $-2.1\times 10^{-4}$ & $-7.5\times 10^{-5}$ &$-3.5\times 10^{-5}$\\

 1800 & $1.2\times 10$ & $1.1\times 10$ & $1.8\times 10$ & $-2.7\times 10^{-4}$ & $-4.3\times 10^{-4}$ & $-1.6\times 10^{-4}$ \\

 2160 & $2.2\times 10$ & $1.2\times 10$ & $9.3$ & $-2.2\times 10^{-5}$ & $-4.1\times 10^{-4}$ & $-1.2\times 10^{-4}$ \\
  \noalign{\smallskip}\hline
\end{tabular}
\ec
\end{table}

Table 3 shows the maximum and average value of the calculated three components of the electric field and the third image of Figure 5 shows the average of absolute value of electric field at six layers. The average of absolute electric field reaches maximum point at the layer of 360 km above the photosphere.

\begin{figure}[h]
   \centering
  \includegraphics[width=9cm, angle=0]{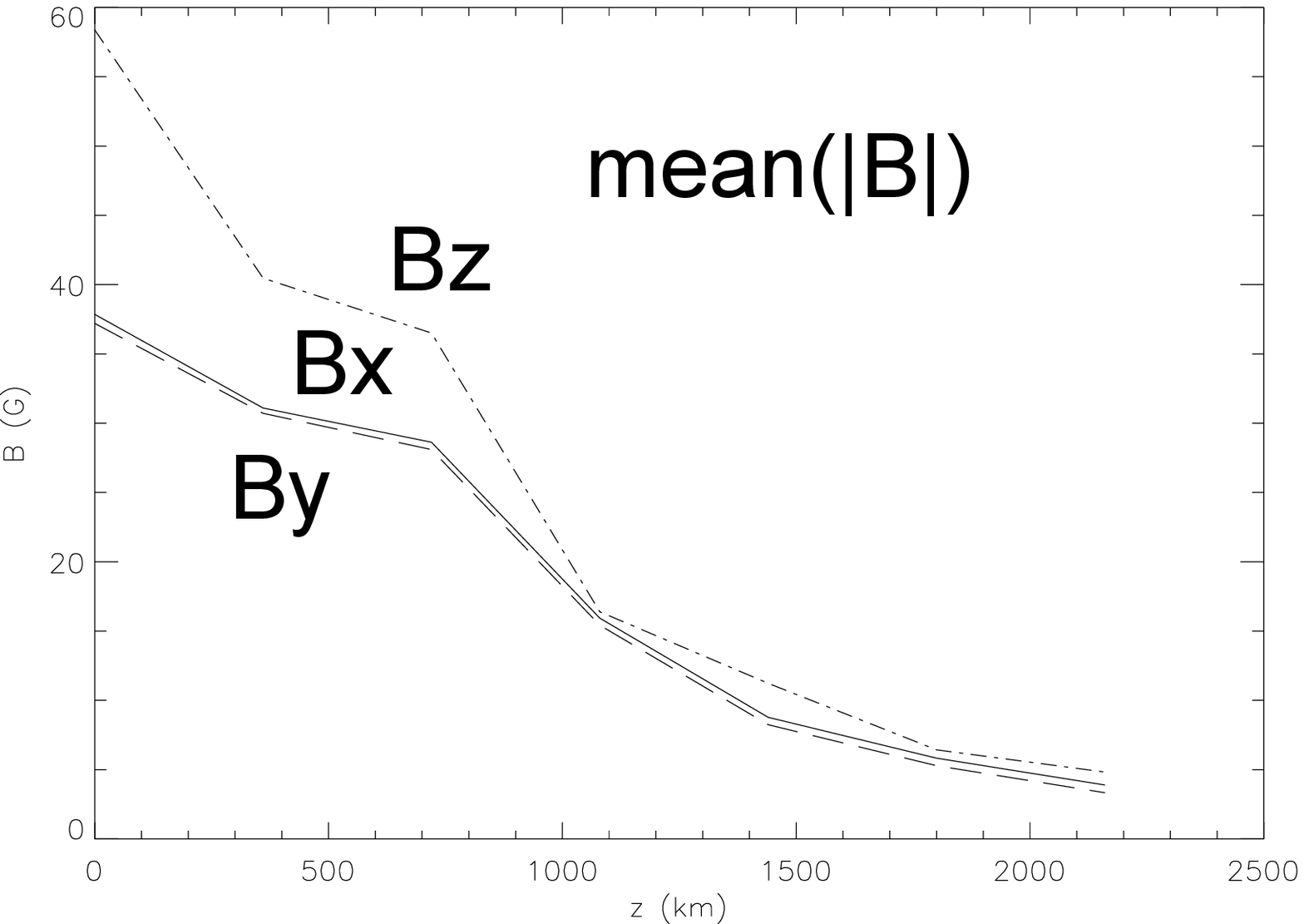}
  \includegraphics[width=9cm, angle=0]{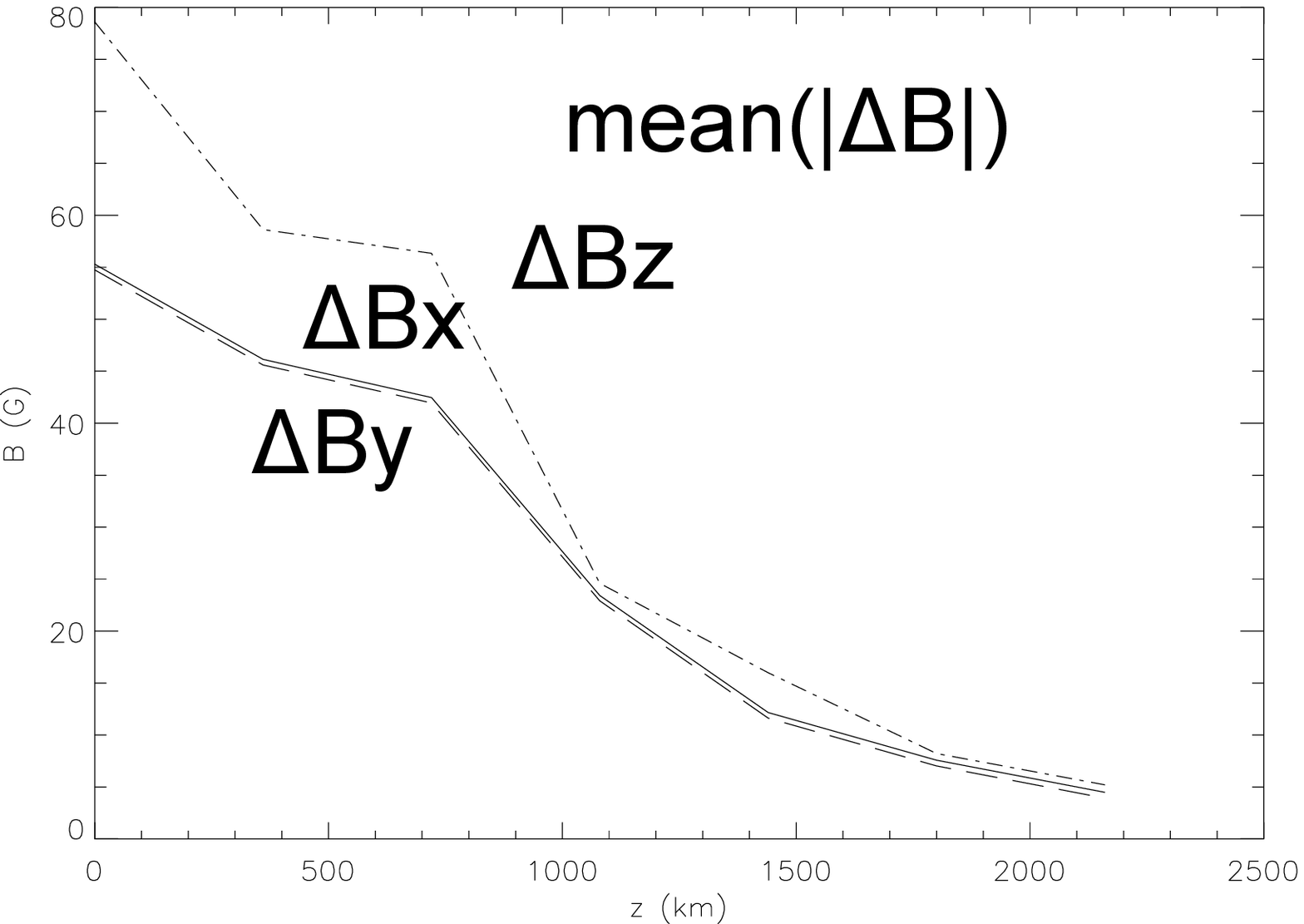}
  \includegraphics[width=9cm, angle=0]{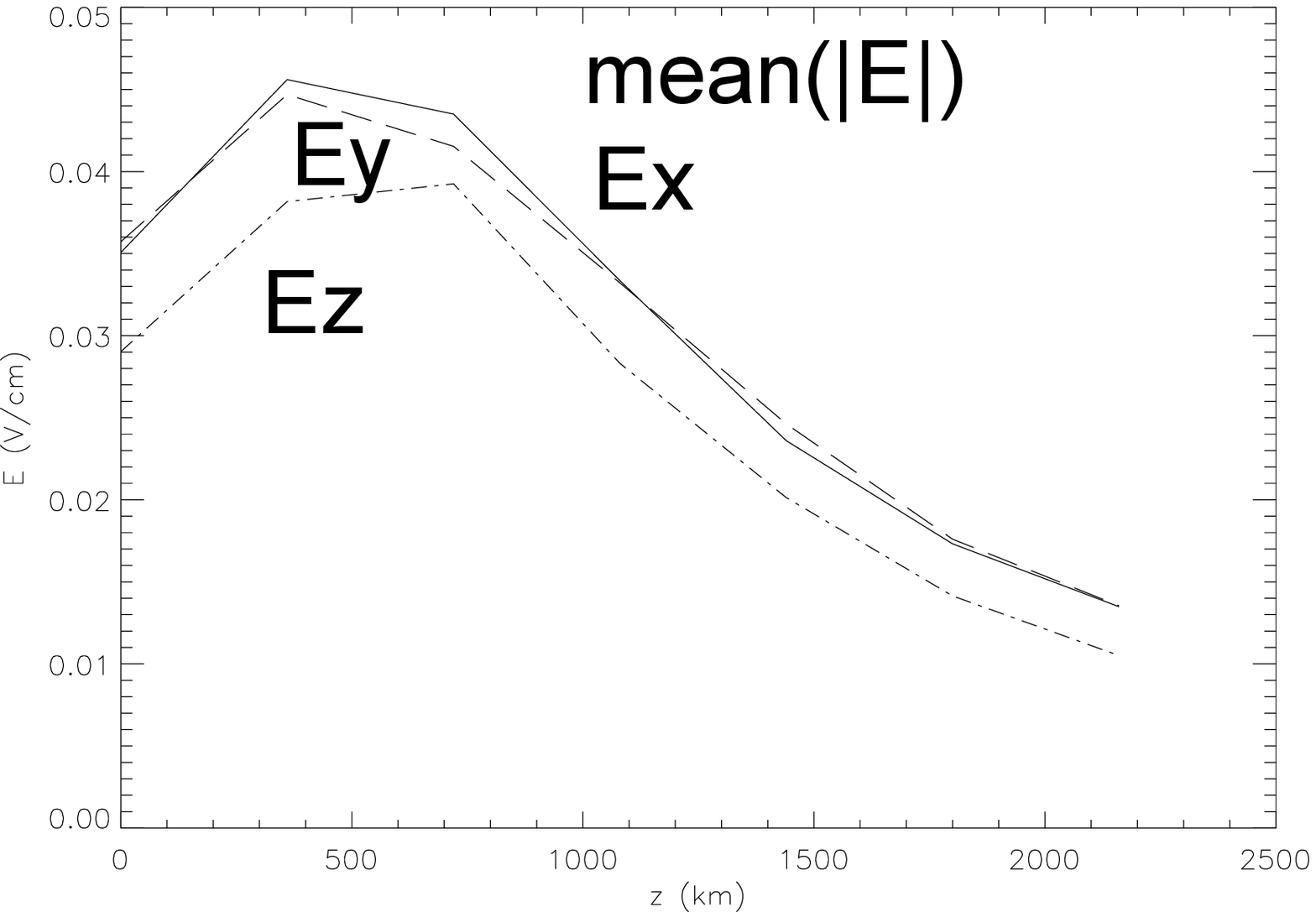}
   \caption{The average of absolute intensity of $\vec{B}$, $\vec{\Delta B}$ and $\vec{E}$ at six layers.}
   \label{Fig5}
   \end{figure}

\section{Summary and Discussion}
\label{sect:discussion}
In this paper, we describe and implement a new method to calculate the distribution of induced electric field in solar atmosphere using a sequence of vector magnetograms as an input.

We first introduce several extrapolation methods of magnetic field and do a simple comparison of these methods, then we choose the optimization method in our example to extrapolate the magnetograms observed by HMI/SDO from photosphere to corona. After that we derived a special solution of electric field in the form of triple integral.

To solve the triple integral problem, we utilize the Monte Carlo statistical method to get a new equation of electric field. As this method require the continuous function, we linearize $\Omega$ which is originally spatial discrete function. The similar linearization has been used to compute time variation of magnetic field from the magnetograms of 12-minute time resolution, that is, we assume that $\vec{B}$ go through linear change in this 12-minute interval.

Through the derivation, it is proved that as long as boundary condition (\citealt{Batchelor+2000}) is fulfilled, we can obtain three-component electric field of solar atmosphere only from vector magnetograms. In our example for NOAA AR11072, the result shows that intensity distribution of induced electric field varies at different layer: it reaches a value of $10^2$ V/cm and the average electric field has a maximum point at the layer of 360 km above the photosphere.
l
However, there are several shortcomings of this method needed to be figured out. At first, the boundary condition (\citealt{Batchelor+2000}) is not strictly satisfied, because in the quiet area, time variation of magnetic field is small but not zero. Secondly, the temporal and spatial resolution of the HMI/SDO vector magnetogram is still not high enough to provide continiuous time and spatial sequences, linearlization methods have to be used twice in our calculation: one is to compute $\partial \vec{B}/ \partial t$ which assume magnetic field linearly change at this 720s interval, and the other is to satisfy the requirement of Monte Carlo method which needs spatial continuous function as input. Thus our calculation process might not fully reflect the real situation of the sun. Finally, we use Monte Carlo method to compute the triple integration of a large volume which is a time-consuming job, and we have to adapt a small sample of only 10,000 random numbers which is far away from large enough.

\normalem
\begin{acknowledgements}
We thank the HMI science team for providing the pre-processecd vector magnetograms. We are very grateful to the anonymous referee whose comments and suggestions help us to improve our manuscript quite a lot. This work is supported by the grant of the key laboratory of solar activity of National Astronomical Observatories, Chinese Academy of Sciences. This work is also supported by the National Natural Science Foundations of China (U1231104, 10921303, 11178005 and 11203036).
\end{acknowledgements}

\bibliographystyle{raa}
\bibliography{bibtex}

\end{document}